\documentclass[epj]{svjour}
\usepackage{graphics}
\newcommand{\cc}{\cite}

\newcommand{\be}{\begin{equation}}
\newcommand{\ee}{\end{equation}}
 
\def\ort{\hbox{$\cal T$}}
\def\lagr{\hbox{$\cal L$}}
\def\ve{\varepsilon}

\def\pd{\partial}

\def\L{\Lambda}

\def\ex{\hbox{e}}

\def\<{\langle}
\def\>{\rangle}

\def\ch{\cosh}

\def\a{\alpha}
\def\b{\beta}
\def\g{\gamma}  \def\G{\Gamma}
\def\d{\delta}  
\def\l{\lambda}   \def\L{\Lambda}
\def\s{\sigma}
\def\r{\rho}  
\def\x{\xi}
\def\c{\chi}
\def\m{\mu}
\def\n{\nu}
\def\t{\tau}

\def\tt{\theta}

\def\({\left(}
\def\[{\left[}
\def\){\right)}
\def\]{\right]}
\def\coth{\hbox{coth}}

\def\pd{\partial}
\def\dk{{d^n k \over (2\pi)^n}}

\def\pa{{\cal P}}

\begin{document}
\title{Soft gluon resummation \protect \\ in quark-vector boson vertex at high energy}
\subtitle{On-mass-shell case}
\author{Igor O. Cherednikov\inst{1,2}\thanks{e-mail: igor.cherednikov@jinr.ru}%
}                     % Do not remove
\institute{Dipartimento di Fisica, Universit\`a di Cagliari and INFN Sezione di Cagliari \\
C.P. 170, I-90142 Monserrato (CA), Italy \and Joint Institute for Nuclear Research\\
BLTP JINR, RU-141980, Dubna, Russian Federation}
\date{Received: 30 November 2004 / }
\abstract{
Resummation of the soft gluon radiative corrections for the
quark-vector boson vertex is performed within the path-integral
(world-line) approach. The leading order expression for the vacuum
averaged Wilson integral for an arbitrary gauge field is found in
$n$-dimensional space-time. The cusp anomalous dimension of the
color non-singlet Sudakov form factor of {\it on}-mass-shell quark
is calculated in an arbitrary covariant gauge in the one-loop
order, and the leading double-logarithmic asymptotical behavior is
obtained from the corresponding evolution equation.
 % end of PACS codes
} %end of abstract
\maketitle
\section{Introduction}
\label{intro}
Resummation of the logarithmic corrections in high-energy
processes is required to justify effectiveness of the Standard
Model at highest energies accessible at modern and future
colliders (see, {\it e.g.}, \cc{COLL} and references therein). In
electroweak processes, the Sudakov corrections can influence
profoundly the cross sections at $e^+ e^-$ linear colliders at
$TeV$ energies, and the precise evaluation of them is quite
important in search for the New Physics, as well as to test the
predictive power of the Standard Model \cc{EW}. In the strong
interactions sector, the form factors of quarks are the most
elementary entities exhibiting the double-logarithmic behavior.
The quark form factors being, of course, unobservable quantities,
enter into the quark-photon and quark-gluon vertices in the
calculations of various QCD processes at the partonic level, and
are of a special theoretical as well as phenomenological
importance \cc{HARD,FEN}. Very recently, the investigations of the
form factors of quarks has been connected with the progress in
experimental study of the constituent quarks and search for the
perturbative and nonperturbative effects in their structure from
low to high energy \cc{SIM,KOCH,OUR}.

The form factors of the elementary fermions at large momenta
transfers in gauge theories---in QED, and later in QCD---had been
studied extensively since the fifties \cc{SUD,KUR,HIST}
\cc{NSOLD,SITYF}. In the color singlet case---corresponding to the
elastic {\it on}-shell quark scattering in an external
electromagnetic field---the exponentiation of the infrared
singularities has been proved and the correct leading asymptotic
at high energy has been obtained \cc{HIST,WREN,KRON,STEF}: the
high-energy behavior of the on-mass-shell quark form factor can be
described in terms of the perturbative evolution equation
\begin{eqnarray} & & F^{qq\g}\[Q^2\]/ F^{qq\g}\[Q_0^2\]  = \nonumber \\  & & \exp\(-\int_{Q_0^2}^{Q^2}\!
\frac{d\m}{2\m} \ \ln\frac{Q^2}{\m}\ \G_{cusp}\[\a_s(\m)\] +
NLO \) \ . \label{ee_em} \end{eqnarray} In
this case, the so-called cusp anomalous dimension reads in
one-loop order \cc{KRON} \be \G^{qq\g}_{cusp}(\a_s) =
\frac{\a_s}{\pi} C_F + O(\a_s^2) > 0 \ , \label{ad_em} \ee and,
therefore, the IR sensitive part of the electromagnetic quark form
factor experiences the Sudakov suppression at large $Q^2$. On the
other hand, it has been shown recently, that in the similar
electroweak reactions the cusp anomalous dimension has the
opposite sign for charged bosons $W^{\pm}$, and then the
enhancement takes place instead of the suppression \cc{ERM}.

Here we apply the powerful Wilson integrals techniques
\cc{NACH,KRON,KKS,STEF} to study of the Sudakov resummation of
soft gluon radiative corrections to the quark-vector boson vertex
with large transferred momentum. In particular, we concentrate on
the $qqg$-vertex, since $qq\g$-vertex can be trivially restored
from the latter. The world-line-based formulations of quantum
theories are actively developed not only due to the wide range of
applications to the standard perturbative QFT, but also from the
point of view of various string theories (for recent review, see
\cc{STR} and Refs. therein).

The very important feature of this approach is that it does not
refer directly to the standard perturbative techniques what allows
one to avoid diagrammatic calculations which use to be very
difficult and involved in non-Abelian gauge theories. Therefore,
it is equally suitable for the perturbative as well as
nonperturbative calculations \cc{ST}. The nonperturbative
calculations within the the world-lines approach could be compared
with the results obtained in lattice QCD and other nonperturbative
frameworks. The quark-vector boson vertex, being one of the
fundamental objects in the theory of strong interactions, is under
active investigation nowadays. The resummation of the gluon
radiative corrections to the $qqg$-vertex has a direct relation to
the problem of IR behavior, chiral symmetry breaking, and
confinement in QCD (see, Ref. \cc{QGV} and Refs. therein). In
high-energy scattering, it can contribute, {\it e.g.,} to the
amplitude with two-gluon exchange in the $t$-channel in the color
singlet state.

The RG equation and corresponding gauge-dependent cusp anomalous dimension
which defines the IR properties of the non-singlet quark form factor
were studied for the first time in Ref. \cc{SITYF} in the leading logarithmic approximation
within the standard diagrammatic approach. In the present paper,
we make an attempt to generalize the path-integral formulation
to colored quark-vector boson vertices and found in one-loop
order the cusp anomalous dimension of the averaged path-ordered
Wilson exponential corresponding to the color non-singlet quark
form factor. We point out an erroneous sign of one of the
terms of the anomalous dimension calculated in \cc{SITYF},
and find and check the correct expression.
Thus, by virtue of exponentiation, one concludes that the colored quark form factor
demonstrates the enhancement at large $Q^2$, in contrast to the
colorless case, Eq. (\ref{ee_em}). The origin of this enhancement
is in close connection with the structure of the gauge group and
has the same nature as the similar behavior in the decays of the
charged electroweak bosons into fermions \cc{ERM}.

\section{Soft gluon resummation within the world-line approach}
\label{sec:1}

In this Section, we generalize the world-line method for
evaluation of the QCD amplitudes with resummed radiative gluon
corrections developed in \cc{STEF} to the case of the quark-vector
boson vertex. For this purpose, let us consider the color
non-singlet form factor of a quark which can be extracted from the
amplitude of the quasi-elastic (color changing) quark scattering
(in the given kinematics) in an external colored (gluon) gauge
field: the quark on a mass shell comes from infinity, emits a hard
gluon at the origin, changes the color, and goes away to infinity:
\begin{eqnarray} & &  u_i(p_1)
\[{\cal M}^{qqV}_\m\]^a_{ij} v_j(p_2) = \nonumber \\ & & F^{qqV}\[(p_1-p_2)^2 ; \x
\] \bar u_i(p_1) T^a_{ij}\g_\m v_j(p_2) \ \ , \label{eq:ampl}
\end{eqnarray} where $p_{1,2}$ are the momenta of the in-coming and out-going
quarks, and $a, (i,j)$ are the gluon and quark color indices,
respectively. This vertex is gauge non-invariant, therefore the
dependence from the gauge parameter $\x$ is included. We consider
the covariant gauge with the following gauge-fixing term in the
Lagrangian: \be \lagr_{gf}^{COV} = - \frac{\l}{2}\sum_a (\pd
A^a)^2 \ \ , \ \ \x = 1 - \frac{1}{\l} \ \ , \label{eq:gf} \ee
while the other important case---the axial gauge $\lagr_{gf}^{AX}
= - \l'/2 (n \cdot A)^2$---deserves a special attention. For the
choice of kinematics with the large (longitudinal) momentum
transfer between quarks (Sudakov regime), the interactions of
quarks with external gluon, as well as the (external-)gluon-gluon
and gluon-ghost interactions were shown to be IR-safe, thus they
can be neglected in the considered case \cc{NSOLD}. The kinematics
is fixed by the small masses of the quarks and large squared
transferred momentum: \be m^2 = p_1^2 = p_2^2 \ \ , \ \ (p_1p_2) =
m^2 \ch \c \ , \ (p_1p_2) \gg m^2 \ee or
\begin{eqnarray} & & s = (p_1 + p_2 )^2 =
2m^2 (1 + \ch \c) \ , \nonumber \\ & & \ - Q^2 = t = (p_1 -p_2 )^2 = 2m^2 (1 - \ch
\c) \ , \nonumber \\  & & \ (s+t) \gg (s-t) \ .
\end{eqnarray} According to the Feynman
rules, the color structure of the vertex function (\ref{eq:ampl})
is determined by the matrix $t_{ij}^a$ in the fundamental
representation of the Lie algebra of the color gauge group
$SU(N_c)$.

Within the world-line formalism, the two-point (Euclidean)
fermionic Green function can be written in terms of the Polyakov
path-integral representation \cc{KKS,STEF}: \begin{eqnarray} & & G_{ij} (x,y) = \nonumber
\\ & & -i \int\! d\t \ \ex^{-m^2 \t} \ \int_{x_0 = x}^{x_\t = y}\! {\cal
D}x(\t') \[m - \frac{1}{2} \g\cdot \dot x (\t)\] \cdot \nonumber \\ & & \pa
\exp\(\frac{i}{4} \int_0^\t\! d\t' \ \s_{\m\n} \omega_{\m\n} \) \cdot
\exp\(-\frac{1}{4} \int_0^\t\! d\t' \ \dot x^2(\t')\) \cdot \nonumber \\
& & \cdot \pa \exp\[ig \int_0^\t\! dx_\r \hat A_\r (x(\t')) \
\]_{ij} \ \ .
\end{eqnarray}
Here $\omega_{\m\n} = \frac{\t}{2}(\ddot x_\m \dot x_\n -
\dot x_\m \ddot x_\n)$ is the Polyakov spin factor, $\s_{\m\n} =
[\g_\m,\g_\n]$, $(i,j)$ are the quarks color indices, and $\t'$ is
treated as the proper time of a fermion with mass $m$ traveling
from the point $x$ at $\t'=0$ to $y$ at $\t'=\t$. Now we are
interesting in the process where a fermion with initial momentum
$p_1$ and color $i$ starts its way at point $x$, passes through a
point $z$ where it experiences both the large momentum transfer $Q
= p_1 - p_2 \ , \ Q^2 < 0 $ and the color changing described by
the matrix $T^a_{ij}$, and comes finally to point $y$ with the
momentum $p_2$ and color $j$. This process can be studied in terms
of the three-point vertex function \be V^\m_{ij} (x,y,z) =
G_{ii'}(x,z) \G^{\m}_{i'j'} G_{j'j}(z,y) \ , \ee where the vertex
matrix $\G^{\m}_{kl}$ consists of a Dirac component $\G^\m =
\g^\m, \g_5\g^\m ...$ and a color component $T^a_{kl} = \d_{kl},
t^a_{kl}$: $\G^{\m}_{kl} = \G^\m \otimes T^a_{kl}$. For
simplicity, we will consider the last case which corresponds to
the quark-gluon vertex, while the color singlet quark-photon
result is restored easily. Therefore, in the momentum space, the
vertex function reads \begin{eqnarray} & &  \tilde {\[ V^\m_{ij} \]}^a (p_1,p_2) = \nonumber \\
& &  \sum_{C_k^{(1)}} \sum_{C_l^{(2)}} \tilde \G^\m \[C_{k,l}\]
\cdot \ort \Bigg\{  \pa \exp\[ ig \int_{C_k^{(1)}} \! dx_\m \ \hat
A_\m(x) \] \cdot \nonumber \\ & &  \cdot T^a_{ij} \cdot  \pa \exp\[ ig \int_{C_l^{(2)}} \! dx_\m \
\hat A_\m(x) \] \Bigg\} \ , \label{eq:vert1} \end{eqnarray} where the sum
over all possible trajectories $C^{(1,2)}_{k,l}$ of two quarks is
assumed. The functions $\G^\m \[C_{k,l}\]$ accumulate information
about quark propagators. In this language, the extraction of the
soft (long-distance) part can be performed by choosing a special
set of paths with simple geometry for ordered exponentials in Eq.
(\ref{eq:vert1}). In our case, one should take for this purpose an
angle with semi-infinite sides that represent the classical
trajectories of the quarks \cc{KRON,STEF}. Then, the UV cutoff
which arises in evaluation of the ``soft'' integrals is identified
with the IR cutoff of the ``hard'' part. After renormalization of
the soft exponentials (for details, see Refs. \cc{WREN}), one can
find the so-called cusp anomalous dimension which determines the
large-$Q^2$ asymptotic of the form factor \begin{eqnarray} & &  F^{qqV}\[Q^2\] = \nonumber \\
& & \exp\(-\int_{Q_0^2}^{Q^2}\! \frac{d\m}{2\m} \ \ln\frac{Q^2}{\m}\
\G_{cusp}^{qqV}\[\a_s(\m)\] + NLO\) \ . \label{eq:evol} \end{eqnarray}
By virtue of the Eq. (\ref{eq:vert1}), it is convenient to express
the latter in terms of the vacuum average of the two path-ordered
Wilson exponentials: \be T_{ij}^a \ F^{qqV}
\[Q^2; \x\] = \Big<0\Big|\ort \Big\{ W_{ii'} T^a_{i'j'} W_{j'j}
\Big\}\Big|0\Big>  \ . \label{eq:def1} \ee Taking into account
that Eq. (\ref{eq:def1}) contains, in general, both UV and IR
divergences, we use scales $\m^2$ and $\l^2$ as the UV and IR
regulator, respectively. For the on-mass-shell quarks, their
trajectories are the semi-infinitely extended paths, and can be
parameterized as: \be \hbox{{\bf In}:} \ \ x_\m =
v^1_\m \ \t \ \ , \ \ \t \in [-\infty, 0] \ \ , \ \ v^1_\m =
p^1_\m / m \ , \ee \be \hbox{{\bf Out}:} \ \ y_\n =
v^2_\n \ \s \ \ , \ \ \s \in [0, +\infty] \ \ , \ \ v^2_\n =
p^2_\n / m \ . \ee Thus, the path-ordered exponentials can be
written as \be W_{ii'} = \pa \exp
\[ig \ t^\a v^1_\m \int_{-\infty}^0 \! d\s \ A^\a_\m (v^1 \s) \]
\Bigg|_{ii'} \ee and \be W_{j'j} = \pa \exp \[ig \ t^\b v^2_\m
\int_{0}^\infty \! d\s \ A^\b_\m (v^2 \s) \]\Bigg|_{j'j} \ee The
non-zero contributions (up to $O(g^2)$ terms) to $F^{qqg}$
(\ref{eq:def1}) stem from the terms: \begin{eqnarray} & &  W_0 = \d_{ii'} \d_{j'j}
t^a_{i'j'} = t^a_{ij} \ , \label{w0} \end{eqnarray}
\begin{eqnarray} & &  W^{(1)}_{LO} =  - {g^2 \over 2}\ t^a_{ij} \ C_F \ v^1_\m v^1_{\m'}
\cdot \nonumber \\ & &  \int_{0}^\infty \!
d\s \int_0^\infty \! d\s' \ \tt(\s-\s') \ D_{\m\m'}
\[v^1(\s-\s')\] \ , \label{eq:qdep0} \\ & &  W^{(2)}_{LO} =
- {g^2 \over 2}\ t^a_{ij} \ C_F \ v^2_\m v^2_{\m'} \cdot \nonumber \\ & &  \int_{-\infty}^0 \!
d\s \int_{-\infty}^0 \! d\s' \ \tt(\s-\s') \ D_{\m\m'}
\[v^2(\s-\s')\] \ , \\  & &  W^{(12)}_{LO} = - {g^2 \over 2}
\(C_F - \frac{C_A}{2}\) \  t^a_{ij}\ v^1_\m v^2_\n\ \cdot \nonumber \\
& & \int_{0}^\infty \! d\t \int_0^\infty \! d\s \ D_{\m\n}
(v^1\t+v^2\s) \ . \label{eq:qdep1} \end{eqnarray} Here the free gluon
propagator is \be
 \Big<0\Big|\ort A^\a_\m (x) A^\b_\n (y)\Big|0\Big> =
 {\cal D}_{\m\n}^{\a\b}(x-y) = \d^{\a\b} D_{\m\n}(x-y)  \ ,
\ee and the following relations have been used: \begin{eqnarray} & &  t^\a_{ki}
t^\a_{il} =  {N_c^2 -1 \over 2N_c} \d_{kl} = C_F \d_{kl} \ \ , \nonumber \\
& & t^\a_{ij} t^\a_{kl} = {1 \over 2} \[\d_{il}\d_{jk} - {1 \over N_c}
\d_{ij}\d_{kl}\] \ .
\end{eqnarray}

\section{Leading order contributions for an arbitrary gauge field}

First, we evaluate the vacuum averaged Wilson integral Eq.
(\ref{eq:def1}) in $n$-dimensional space-time, for an arbitrary
gauge field which can be of any origin, for
instance---nonperturbative. Then the general results will be
applied in the case of usual perturbative gluon field.

It is convenient to present the gauge field propagator
$D_{\m\n}(z)$ in the form \cc{OUR}: \be D_{\m\n}(z) = g_{\m\n}
\pd_\r \pd^\r D_1(z^2) - \pd_\m\pd_\n D_2(z^2) \ . \label{st1} \ee
Calculation of the integrals (\ref{eq:qdep0}, \ref{eq:qdep1})
requires the following expressions: \be \pd_\r \pd^\r = 2 n \pd +
4 z^2 \pd^2 \ , \ \pd_\m\pd_\n = 2 g_{\m\n} \pd + 4 z_\m z_\n
\pd^2 \ , \ \pd = \pd_{z^2} \ . \ee The scalar products are:
\begin{eqnarray}
& & (v^{1,2}_\m z_\m)^2 = (\s-\s')^2 \ ,  \ v^1_\m v^2_\n
g_{\m\n} = \cosh \c \ , \nonumber \\ & &  \ v^1_\m v^2_\n z_\m z_\n = (v^1\s +
v^2\t)^2 \cosh \c  - \s\t \sinh^2 \c \ .
\end{eqnarray} In order to control
UV singularity, the transverse space-like separation $\vec b =
\(0_\parallel, \vec b_\perp \)$ between two points at the
integration path is introduced: \begin{eqnarray} & & (\s-\s')^2 \to (\s-\s')^2
-\vec b^2 \ , \nonumber \\ & &  (v^1\s + v^2\t)^2 \to (v^1\s + v^2\t)^2 - \vec
b^2 \ .
\end{eqnarray}

The path-ordered integrals from Eqs.
(\ref{eq:qdep0},\ref{eq:qdep1}) can be evaluated using the
following basic integrals: \begin{eqnarray} & & \int_0^{\infty} \! d\s  d\s' \
\ex^{-\a (\s-\s')^2 } = -\frac{1}{2\a} \ , \nonumber \\ & &   \int_0^{\infty} \!
d\s  d\s' \ \s\s' \ \ex^{-\a (\s-\s')^2} = -\frac{1}{12\a^2} \ , \nonumber \\
& &  \int_0^{\infty} \! d\s d\t \ \ex^{-\a (\s^2 + \t^2 + 2
\s\t \ \cosh \c)} = {1 \over 2 \a}\frac{\c}{\sinh \c} \ , \nonumber \\
& & \int_0^{\infty} \! d\s d\t \ \s \t \ \ex^{-\a (\s^2 + \t^2 + 2 \s
\t \cosh \c)} = \nonumber \\ & & = \frac{\c \coth \c -1}{4 \a^2 \sinh^2 \c} \ .
\end{eqnarray}
Then, applying the Laplace transform to the invariant functions
$D_i (u)$, and it's derivatives over $u=z^2$: \be D_i^{\{k\}}(u) =
(-)^k \int_0^{\infty} \! d\a \ \a^k \ \ex^{-\a u} \bar D_i(\a) \ ,
\ee one obtains the following formula:
\begin{eqnarray} & & \int_0^{\infty} \! d\s
d\s' \ D' (u) = \frac{1}{2} D (-\vec b^2) \ , \nonumber \\ & & \int_0^{\infty} \!
d\s d\s' \ D'' (u) = \frac{1}{2} D'(-\vec b^2) \ , \\
& & \int_0^{\infty} \! d\s d\s' \ \s\s' \ D'' (u) = \frac{1}{12}
D(-\vec b^2) \ , \\ & &  \int_0^{\infty} \! d\s d\s' \ (\s-\s')^2
\ D'' (u) = - \frac{1}{2} D(-\vec b^2) \end{eqnarray} for $u = (\s-\s')^2 -
\vec b^2$, and \begin{eqnarray} & & \int_0^{\infty} \! d\s d\t \ D' (u) = -
\frac{\c}{2 \sinh \c} D(- \vec b^2) \ , \nonumber \\ & &  \ \int_0^{\infty}
\! d\s d\t \ \s \t D'' (u) = \frac{\c \coth \c -1}{4 \sinh^2 \c}
D(- \vec b^2) \ , \nonumber \\
& &  \int_0^{\infty} \! d\s d\t \ (v^1\s + v^2\t)^2 D'' (u) = {\c
\over 2 \sinh \c} D(- \vec b^2) \ , \nonumber \\ & &  \int_0^{\infty} \! d\s
d\t \ D'' (u) = \frac{\c}{2 \sinh \c} D'(- \vec b^2) \
\end{eqnarray} for $u = (v^1\s + v^2\t)^2 - \vec b^2$.

Thus, one finds the general expressions in $n$ dimensions and
arbitrary covariant gauge, with the gauge field two-point
correlator expressed like the Eq. (\ref{st1}): \begin{eqnarray} & & W^{(1)} = W^{(2)} =
- t^a_{ij}\ \frac{g^2}{2} C_F \cdot \nonumber \\ & & \cdot \[(n-2) D_1(-\vec b^2) + 2 \vec b^2
D'_1(-\vec b^2) + D_2 (-\vec b^2) \] \ , \label{gen1} \end{eqnarray}
\begin{eqnarray} & & W^{(12)}(\c) =  t^a_{ij}\  g^2 \(C_F - \frac{C_A}{2} \) \cdot \nonumber \\
& & \cdot \[\c\coth \c \((n-2) D_1(-\vec b^2) + \right. \right. \nonumber \\
& & \left. \left. 2 \vec b^2 D'_1(-\vec b^2)
\) + D_2(-\vec b^2)\]  \label{gen2} \end{eqnarray} up to $O(g^4)$ order
terms. Let us emphasize, that the equations (\ref{gen1},
\ref{gen2}) derived above are valid for any gauge field in the
adjoint representation of the $SU(N_c)$ color group. Therefore,
these results can be used for evaluation of the nonperturbative,
{\it e.g.,} instanton, contributions to the vacuum averaged Wilson
integrals in the Eq. (\ref{eq:def1}).

\section{One-loop perturbative contribution}

Now let us apply this general result in the particular case of the
perturbative gluon field. Here and in what follows, the
dimensional regularization is used with $n= 4 - 2\ve$, $\ve < 0$
in order to regulate the IR-divergent terms in the integrals,
respecting, in the same time, the gauge invariance. The gluon
propagator in the coordinate space reads:
\begin{eqnarray} & & D_{\m\n}(z; \x) =  \nonumber \\ & & \frac{\l^{4-n}}{i}
\int \! \dk \frac{\ex^{-ikz}}{ k^2+i0 }
\[g_{\m\n} - \x \frac{k_\m k_\n}{k^2+i0} \] = \frac{1}{4\pi^2}\(-\pi\l^2\)^{\ve} \cdot \nonumber \\
& & \[g_{\m\n}
\frac{\G(1-\ve)}{\(z^2 -i0\)^{1-\ve}} + \x \pd_\m\pd_\n \ \frac{
\G(-\ve) }{ \(z^2 -i0\)^{-\ve} } \] . \label{eq:prop1} \end{eqnarray} The
perturbative one-loop invariant functions in Eq. (\ref{st1}) then
read: \begin{eqnarray} & & D_1(z^2) = \frac{1}{16\pi^2} \frac{\G(1-\ve)}{\ve}
\(-\pi\l^2z^2\)^{\ve} \ , \nonumber \\ & &  D_2(z^2) = - \x D_1(z^2) \ .
\end{eqnarray}
Now the Eqs. (\ref{gen1}, \ref{gen2}) contain the IR singularities at
$\ve \to 0$. Performing the standard renormalization procedure
within the ${\overline{MS}}$ scheme, described in detail in Refs.
\cc{WREN,MMP,KRON}, one finds \begin{eqnarray} & & W_{LO}^{(1)}(\a_s, \m^2/\l^2;
\x) = W_{LO}^{(2)}(\a_s, \m^2/\l^2; \x) = \nonumber \\ & & t^a_{ij}
\frac{\a_s}{4\pi} C_F \( 1 - \frac{\x}{2} \)\ \ln
\frac{\m^2}{\l^2}  \ , \label{w1} \end{eqnarray} and \begin{eqnarray} & & W_{LO}^{(12)} (\a_s,
\c, \m^2/\l^2; \x) = \nonumber \\ & & - t^a_{ij} \frac{\a_s}{2\pi} \(C_F -
\frac{C_A}{2} \) \ \[ \c \coth \c - \frac{\x}{2}
\] \ln \frac{\m^2}{\l^2} \ . \label{w12} \end{eqnarray} Here the
UV-normalization point is taken $\m^2 = 4 \vec b^{-2}$.

Then one needs to combine the expression for the one-loop
contribution to the form factor from Eqs. (\ref{w0}, \ref{w1},
\ref{w12}): \begin{eqnarray} & & F^{qqV}_{LO} (Q^2; \x) =  W_0 + 2 W_{LO}^{(1)}
\(\a_s, \frac{\m^2}{\l^2}\) + \nonumber \\ & & + W_{LO}^{(12)}\(\a_s,
\frac{\m^2}{\l^2}, \c \) + O(\a_s^2)\ .
\end{eqnarray}

The high-energy asymptotic behavior of the form factor is
determined by the (gauge-dependent) cusp anomalous dimension which
stems from the renormalization of the Wilson integral
(\ref{eq:def1}) \cc{NSOLD,SITYF,KRON}: \begin{eqnarray} & & \(\m^2 \frac{\pd}{\pd \m^2} +
\b(\a_s)\frac{\pd}{\pd \a_s} + \d (\a_s, \x) \x \frac{\pd}{\pd \x}
\) \ln F^{NS} (Q^2) = \nonumber \\ & & - \frac{1}{2} \G_{cusp}^{qqg}
\[\a_s(\m^2); \c \] \ , \label{ad} \end{eqnarray} with the one-loop
functions \begin{eqnarray} & & \b(\a_s) = \m^2 {\pd \a_s \over \pd \m^2} = \nonumber \\ &  &
- \frac{\b_0}{4\pi} \a_s^2\ + O(\a_s^3), \ \b_0 = \frac{11}{3}N_c -
\frac{2}{3}n_f \ ,
\end{eqnarray} and (see Ref. \cc{VLAD}) \begin{eqnarray} & & \d(\a_s; \x) =
\m^2 {\pd \ln \x \over \pd \m^2} = \nonumber \\ & & \frac{\a_s}{4\pi} C_A
\frac{\x-1}{\x} \[\frac{\x-1}{2} + \frac{13}{6} \] + O(\a_s^2) \  .
\end{eqnarray} Thus, the calculation of the cusp anomalous dimension
(\ref{ad}) yields in one-loop approximation: \begin{eqnarray} & & \G_{cusp}^{qqg}
\[\a_s(\m^2); \c ; \x \] = \nonumber \\ & & \frac{\a_s}{\pi} \[ \(C_F - \frac{C_A}{2}\)
\c \coth \c + \frac{C_A}{4} \x - C_F  \] + O(\a_s^2)
\label{mainres} \end{eqnarray} where the Casimir operator of $SU(N_c)$ in the
adjoint representation is used: $C_A = N_c$. Note, that in the Ref. \cc{SITYF}, the term
$C_A\x/4$ in Eq. (\ref{mainres}) had the sign {\it minus}. However, our expression
allows to reproduce the color singlet case by means of the replacement of the color
factors: $(C_F - C_A/2) \to
C_F$, thus obtainig the gauge invariant result (\ref{ad_em}):
\begin{eqnarray} & & \G_{cusp}^{qq\g}
\[\a_s(\m); \c \] = \nonumber \\ & & \frac{\a_s}{\pi} \[ C_F
\c \coth \c +  \(C_F - C_F \)\x - C_F  \] + O(\a_s^2) = \nonumber \\ & &
\frac{\a_s}{\pi} \frac{N_c^2 -1}{2N_c} \(  \c \coth \c -1 \) +
O(\a_s^2)\ , \label{check}
\end{eqnarray} yielding the well known Sudakov suppression. The
latter can be considered as a convenient test of the calculations and confirms
the accuracy of our result, Eq. (\ref{mainres}).

In order to find the large-$Q^2$ asymptotic, we take into account
the limit of the large scattering angle: \be \c \coth \c \propto
\ln \frac{Q^2}{m^2} \ \ , \ \ \c \to \infty \ , \ee and find that,
in this limit, the anomalous dimension is  linear in $\ln Q^2$
while the gauge-dependent term can be neglected:
\begin{eqnarray} & &
\G_{cusp}^{qqg} \[\a_s; \c \] =  \G_{cusp}^{qqg} \[
\a_s\] \ln \frac{Q^2}{m^2} + O(\ln^0 Q^2) \ , \\
& & \G_{cusp}^{qqg} \[\a_s\] = \frac{\a_s}{\pi} \(C_F - \frac{C_A}{2}\)
 + O(\a_s^2) < 0 \ . \label{adlim} \end{eqnarray} Note that for the color group $SU(3)$, this anomalous
dimension (\ref{adlim}) has the opposite sign compared to the
color singlet case Eq. (\ref{ad_em}): this is a direct consequence
of the algebra of gauge group generators. Therefore, the leading
(double-logarithmic) behavior of the non-singlet form factor is
given by
\begin{eqnarray} & &
F^{qqg}\[Q^2\]/F^{qqg} \[Q_0^2 \] =  \nonumber \\ & & \exp\[-\(C_F -
\frac{C_A}{2}\)\int_{Q_0^2}^{Q^2}\! \frac{d\m}{2\m} \
\ln\frac{Q^2}{\m}\ \frac{\a_s(\m)}{\pi}  + NLO \]
= \nonumber \\ & &  \exp\[\frac{2}{\b_0 N_c} \ln \frac{Q^2}{\L_{QCD}^2} \ln\frac{\ln
Q^2/\L_{QCD}^2}{\ln Q_0^2/\L_{QCD}^2} + NLO\]  \ , \label{evolfin} \end{eqnarray} that is the increasing
function of $Q^2$. The dependence from the gauge-fixing parameter
$\x$ drops out of the leading logarithmic expression for
$\G_{cusp}^{qqg}$, Eq. (\ref{adlim}), and yields no influence on
the main asymptotics.

\section{Discussion and Conclusions}

We derived the general formula for the vacuum averaged Wilson
integral (\ref{eq:def1}) in the $g^2$ accuracy. This result Eqs.
(\ref{gen1}, \ref{gen2}) is found in $n$ dimensions, in an
arbitrary covariant gauge, and is valid as well for any $SU(N_c)$
gauge field. In the case of a nonperturbative field $A_{NP}^\m
\sim g^{-1} $, this result corresponds to the leading (so-called
``weak field'') order of expansion in field strength, while the
dependence on the coupling $g$ drops out.

By using the Wilson integrals techniques, it has been found that
the cusp anomalous dimension for the non-singlet quark form factor
Eq. (\ref{adlim}) has a negative sign---in contrast to the singlet
case---what leads, in the large-$Q^2$ regime, to the enhancement,
rather than suppression, of the contribution due to the resumed
soft gluon radiative corrections, Eq. (\ref{evolfin}). The
dependence from the gauge-fixing parameter $\x$ is shown to be of
the order of $\ln^0 Q^2$---not surviving in the asymptotical
regime.

It is necessary to emphasize that we work here in the covariant gauge,
whereas the case of the axial gauge is quite important and
requires additional analysis (see, {\it e.g.,} Refs. \cc{AXI}).
Another important version of the problem---the case of the {\it off}-shell quarks with different
masses (applicable, {\it e.g.,} to flavor changing processes) is
more technically involved and will be reported in the forthcoming
work.

Another point which should be noted is that the quark-gluon vertex
is a colored gauge-dependent object and could not be an observable
quantity. In computations of the realistic processes when this
vertex is inserted into the diagrams at the partonic level, the
role of this enhancement can be reduced due to the emission of
gluons, as well as due to the convolutions of the color indices.

\section*{Acknowledgments}
\noindent The author thanks A.E. Dorokhov, L. Magnea and N.G.
Stefanis for careful reading of the manuscript, fruitful
discussions and valuable remarks. The conversations with E.A.
Kuraev and O.V. Teryaev on a number of topics concerning this work
were very useful and inspiring. The work is partially supported by
RFBR (Grant Nos. 04-02-16445, 03-02-17291), Russian Federation
President's Grant 1450-2003-2. The hospitality and financial
support of the INFN Section in Torino (where this work was
started), and the High Energy Section at ICTP in Trieste (where it
was completed) is gratefully acknowledged.


\begin{thebibliography}{99}

\bibitem{COLL}
S. Catani {\it et al.}, $\mathtt{hep-ph/0005025}$;
%%CITATION = HEP-PH 0005025;%%
S. Catani, $\mathtt{hep-ph/0005233}$.
%%CITATION = HEP-PH 0005233;%%

\bibitem{EW}
P. Ciafaloni, D. Comelli, {Phys. Lett.} {\bf B476} (2000) 49;
%%CITATION = HEP-PH 9910278;%%
V.S. Fadin, L.N. Lipatov, A.D. Martin, M. Melles, {Phys. Rev.}
{\bf D61} (2000) 094002;
%%CITATION = HEP-PH 9910338;%%
J.H. Kuhn, A.A. Penin, V.A. Smirnov, {Eur. Phys. J.} {\bf C17}
(2000) 97.
%%CITATION = HEP-PH 9912503;%%




\bibitem{HARD} Yu. Dokshitzer, D. Dyakonov, S. Troyan, {Phys. Reports}
{\bf 58} (1980) 269;
%%CITATION = PRPLC,58,269;%%
A. H. Mueller, {Phys.  Reports}  {\bf 73} (1981) 237.
%%CITATION = PRPLC,73,237;%

\bibitem{FEN} G. Parisi, {Phys. Lett.} {\bf B90} (1980) 295;
%%CITATION = PHLTA,B90,295;%%
G. Curci, M. Greco, {Phys. Lett.} {\bf B92} (1980) 175;
%%CITATION = PHLTA,B92,175;%%
H.n. Li, G. Sterman, {Nucl. Phys.} {\bf B381} (1992) 129;
%%CITATION = NUPHA,B381,129;%%
L. Magnea, G. Sterman, {Phys. Rev.} {\bf D42} (1990) 4222;
%%CITATION = PHRVA,D42,4222;%%
L. Magnea, {Nucl. Phys.} {\bf B593} (2001) 269;
%%CITATION = NUPHA,B593,269;%%


\bibitem{SIM}
R. Petronzio, S. Simula,  G. Ricco, {Phys. Rev.} {\bf D67} (2003)
094004;
%%CITATION = HEP-PH 0301206;%%
S. Simula, {Phys. Lett.} {\bf B574} (2003) 189;
%%CITATION = HEP-PH 0307160;%%
M. Osipenko {\it et al.}  [CLAS Collaboration], {Phys. Rev.} {\bf
D67} (2003) 092001;
%%CITATION = HEP-PH 0301204;%%
M. Osipenko {\it et al.}, $\mathtt{hep-ex/0309052}$.
%%CITATION = HEP-EX 0309052;%%

\bibitem{KOCH} N. Kochelev, {Phys. Lett.} {\bf B565} (2003) 131.
%%CITATION = HEP-PH 0304171;%%

\bibitem{OUR} A. Dorokhov, I. Cherednikov, {Phys. Rev.} {\bf D66} (2002) 074009;
%%CITATION = PHRVA,D66,074009;%%
{Phys. Rev.} {\bf D67} (2003) 114017;
%%CITATION = HEP-PH 0212357;%%
{Annals Phys.} (NY) {\bf 314} (2004) 321;
%%CITATION = HEP-PH 0404040;%%
I. Cherednikov, {Surv. High Energy Phys.} {\bf 18} (2003) 205.
%%CITATION = HEP-PH 0305055;%%

\bibitem{SUD} V. V. Sudakov, {Sov. Phys. JETP} {\bf 3} (1956) 65;
[{Zh. Eksp. Teor. Fiz.}  {\bf 30} (1956) 87].
%%CITATION = SPHJA,3,65;%%

\bibitem{KUR} E. A. Kuraev, V. S. Fadin, {Yad. Fiz.} {\bf 27} (1978) 1107;
%%CITATION = YAFIA,27,1107;%%
B.I. Ermolaev, V.S. Fadin, {JETP Lett.} {\bf 33} (1981) 269;
[{Pisma Zh. Eksp. Teor. Fiz.}  {\bf 33} (1981) 285]; B.I.
Ermolaev, S.I. Troyan, {Nucl. Phys.} {\bf B590} (2000) 521.
%%CITATION = HEP-PH 0009345;%%

\bibitem{HIST} R. Jackiw, {Annals Phys.} (NY) {\bf 48} (1968) 292;
%%CITATION = APNYA,48,292;%%
P. M. Fishbane, J. D. Sullivan, {Phys. Rev.} {\bf D4} (1971) 458;
%%CITATION = PHRVA,D4,458;%%
J. M. Cornwall, G. Tiktopoulos, {Phys. Rev.} {\bf D13} (1976)
3370;
%%CITATION = PHRVA,D13,3370;%%
J. Frenkel, J. C. Taylor, {Nucl. Phys.} {\bf B116} (1976) 185;
%%CITATION = NUPHA,B116,185;%%
E. C. Poggio, G. Pollack, {Phys. Lett.} {\bf B71} (1977) 135;
%%CITATION = PHLTA,B71,135;%%
A. H. Mueller, {Phys. Rev.} {\bf D20} (1979) 2037;
%%CITATION = PHRVA,D20,2037;%%
J. C. Collins, {Phys. Rev.} {\bf D22} (1980) 1478;
%%CITATION = PHRVA,D22,1478;%%
A. Sen, {Phys. Rev.} {\bf D24} (1981) 3281.
%%CITATION = PHRVA,D24,3281;%%

\bibitem{NSOLD} J.J. Carazzone, E.C. Poggio, H.R. Quinn,
{Phys. Rev.} {\bf D11} (1975) 2286; [Erratum-ibid. {\bf D12}
(1975) 3368].
%%CITATION = PHRVA,D11,2286;%%

\bibitem{SITYF}
Yu. A. Sitenko, Sov. J. Nucl. Phys. {\bf 27} (1978) 583; [{Yad.
Fiz.} {\bf 27} (1978) 1098 (in Russian)].
%%CITATION = YAFIA,27,1098;%%



\bibitem{NACH} A. Bassetto, M. Ciafaloni, G. Marchesini, {Phys. Reports} {\bf 100} (1983) 201;
%%CITATION = PRPLC,100,201;%%
O. Nachtmann, {Annals Phys.} (N.Y.) {\bf 209} (1991) 436.
%%CITATION = APNYA,209,436;%%

\bibitem{MMP} Yu.M. Makeenko, A.A. Migdal, {Phys. Lett.} {\bf B88} (1979) 135;
%%CITATION = PHLTA,B88,135;%%
{Nucl. Phys.} {\bf B188} (1981) 269;
%%CITATION = NUPHA,B188,269;%%
A. Polyakov, {Phys. Lett.} {\bf B82} (1979) 247;
%%CITATION = PHLTA,B82,247;%%
{Nucl. Phys.} {\bf B164} (1980) 171.
%%CITATION = NUPHA,B164,171;%%

\bibitem{WREN} V. S. Dotsenko, S. N. Vergeles, {Nucl. Phys.} {\bf B169} (1980)
527;
%%CITATION = NUPHA,B169,527;%%
R. A. Brandt, F. Neri, M.-A. Sato, {Phys. Rev.} {\bf D24} (1981)
879;
%%CITATION = PHRVA,D24,879;%%
R. A. Brandt, A. Gocksch, M. A. Sato, F. Neri, {Phys. Rev.} {\bf
D26} (1982) 3611.
%%CITATION = PHRVA,D26,3611;%%
J.G.M. Gatheral, {Phys. Lett.} {\bf B133} (1983) 90;
%%CITATION = PHLTA,B133,90;%%
J. Frenkel, J.C. Taylor, {Nucl. Phys.} {\bf B246} (1984) 231.
%%CITATION = NUPHA,B246,231;%%

\bibitem{KRON} G. Korchemsky, {Phys. Lett.} {\bf B217} (1989) 330;
%%CITATION = PHLTA,B217,330;%%
G. Korchemsky, A. Radyushkin, {Sov. J. Nucl. Phys.} {\bf 45}
(1987) 910;
%%CITATION = YAFIA,45,910;%%
G. Korchemsky, A. Radyushkin, {Nucl. Phys.} {\bf B283} (1987) 342;
%%CITATION = PHRVA,D56,4043;%%
G. Korchemsky, {Phys. Lett.} {\bf B220} (1989) 629.
%%CITATION = PHLTA,B220,629;%%

\bibitem{KKS} A.I. Karanikas, C.N. Ktorides,
{Phys. Rev.} {\bf D52} (1995) 5883;
%%CITATION = PHRVA,D52,5883;%%
A.I. Karanikas, C.N. Ktorides, N.G. Stefanis, {Phys. Rev.} {\bf
D52} (1995) 5898.
%%CITATION = PHRVA,D52,5898;%%

\bibitem{STEF} G.C. Gellas, A.I. Karanikas, C.N. Ktorides, N.G. Stefanis,
{Phys. Lett.} {\bf B412} (1997) 95;
%%CITATION = HEP-PH 9707392;%%
N.G. Stefanis, $\mathtt{hep-ph/9811262}$;
%%CITATION = HEP-PH 9811262;%%
A.I. Karanikas, C.N. Ktorides, N.G. Stefanis, {Eur. Phys. J.} {\bf
C26} (2003) 445.
%%CITATION = HEP-PH 0210042;%%

\bibitem{ERM} B.I. Ermolaev,
$\mathtt{hep-ph/0211302}$;
%%CITATION = HEP-PH 0211302;%%
B.I. Ermolaev, S.M. Oliveira, S.I. Troyan, {Phys. Rev.} {\bf D66}
(2002) 114018;
%%CITATION = HEP-PH 0207159;%%
A.Barroso, B.I. Ermolaev, $\mathtt{hep-th/0112216}$.
%%CITATION = HEP-TH 0112216;%%

\bibitem{STR} C. Schubert,
{Phys. Reptorts}  {\bf 355} (2001) 73.
%%CITATION = HEP-TH 0101036;%%

\bibitem{QGV}
S.J. Stainsby, R.T. Cahill, {Mod. Phys. Lett.} {\bf A9} (1994)
3551;
%%CITATION = HEP-PH 9306313;%%
A.I. Davydychev, P. Osland, L. Saks, {Phys. Rev.} {\bf D63} (2001)
014022;
%%CITATION = HEP-PH 0008171;%%
A. Bender, W. Detmold, C.D. Roberts, A.W. Thomas, {Phys. Rev.}
{\bf C65} (2002) 065203;
%%CITATION = NUCL-TH 0202082;%%
J.I. Skullerud, P.O. Bowman, A. Kizilersu, D.B. Leinweber, A.G.
Williams, JHEP {\bf 0304}, 047 (2003).
%%CITATION = HEP-PH 0303176;%%


\bibitem{ST} Yu.A. Simonov, J.A. Tjon,
{Annals Phys.} {\bf 300} (2002) 54;
%%CITATION = HEP-PH 0205165;%%
A.V. Belitsky, A.S. Gorsky, G.P. Korchemsky, {Nucl. Phys.} {\bf
B667} (2003) 3.
%%CITATION = HEP-TH 0304028;%%


\bibitem{VLAD}
O.V. Tarasov, A.A. Vladimirov, {Sov. J. Nucl. Phys.}  {\bf 25}
(1977) 585; [{Yad. Fiz.}  {\bf 25} (1977) 1104].
%%CITATION = SJNCA,25,585;%%

\bibitem{AXI} G.T. Bodwin, S.J. Brodsky, G.P. Lepage,
Phys.\ Rev.\ Lett.\  {\bf 47} (1981) 1799;
%%CITATION = PRLTA,47,1799;%%
R. Doria, J. Frenkel, J.C. Taylor, Nucl. Phys. {\bf B168} (1980)
93.
%%CITATION = NUPHA,B168,93;%%


\end{thebibliography}
\end{document}